# Accelerating and scaling mentoring strategies to build infrastructure that supports underrepresented groups in STEM

*Authors:*

2030STEM. 2030STEM Inc., NY, NY USA

*Jennifer D. Adams, PhD, 2030STEM Salon Series Editor; University of Calgary, Calgary, Alberta, CAN*

*David J. Asai, Senior Director, Center for the Advancement of Science Leadership and Culture, Howard Hughes Medical Institute, Chevy Chase, Maryland, USA*

*Ruth Cohen, Interim Executive Director and Strategic Advisor, 2030STEM Inc, New York, New York, USA*

*Alonso Delgado, Dept. Evolution, Ecology and Organismal Biology, The Ohio State University, Columbus, Ohio, USA*

*Stephanie Danette Preston, PhD, Associate Dean for Graduate Educational Equity, The Pennsylvania State University, State College, Pennslyvannia, USA*

*Jackie Faherty\*, PhD, American Museum of Natural History; CoFounder 2030STEM, New York, New York, USA, ORCID 0000-0001-6251-0573*

*Mandë Holford\*, PhD, Hunter College; American Museum of Natural History; CoFounder 2030STEM, New York, New York, USA*

*Erich D. Jarvis, PhD, Professor , The Rockefeller University, New York, New York, USA; Investigator, Howard Hughes Medical Institute, Chevy Chase, Maryland, USA*

*Marisela Martinez-Cola, PhD, Associate Professor Sociology, Morehouse College, Atlanta, Georgia, USA*

*Alfred Mays, Director and Chief Strategist for Diversity and Education, Burroughs Wellcome Fund, Apex, North Carolina, USA*

*Louis J. Muglia, MD PhD, Burroughs Wellcome Fund; Department of Pediatrics University of Cincinnati College of Medicine, Division of Human Genetics, Cincinnati Children's Hospital Medical Center, Cincinnati, Ohio, USA*

*Veeshan Narinesingh, NOAA Geophysical Fluid Dynamics Laboratory; Program in Atmospheric and Oceanic Sciences, Princeton University, Princeton, New Jersey, USA*

*Caprice L. Phillips, Department of Astronomy, The Ohio State University, Columbus, Ohio, USA, ORCID 0000-0001-5610-5328*

*Christine Pfund, PhD, Senior Scientist, Wisconsin Center for Education Research, University of Wisconsin-Madison, Madison, Wisconsin, USA*

*Patricia Silveyra, PhD, Department of Environmental and Occupational Health, Indiana University Bloomington School of Public Health, Bloomington, Indiana, USA*

*\*Co-corresponding authors: jfaherty@amnh.org ; mholford@hunter.cuny.edu*

**ABSTRACT**

The vision of 2030STEM is to address systemic barriers in institutional structures and funding mechanisms required to achieve full inclusion in Science, Technology, Engineering, and Mathematics (STEM) and accelerate leadership pathways for individuals from underrepresented populations across STEM sectors. 2030STEM takes a systems-level approach to create a community of practice that affirms diverse cultural identities in STEM.

Accelerated systemic change is needed to achieve parity and representation in the STEM workforce, and mentorship is crucial to retain talent by ensuring those underrepresented in STEM feel that they belong and can thrive. To support the studies and careers of those underrepresented in STEM, we must increase mentors who have received adequate training in the discipline of mentorship, including cross-cultural mentoring, use of evidence-based mentorship tools to improve the outcomes of mentor/mentee relationships, and mentorship at the institutional versus individual level.

This paper is part of a series of white papers based on 2030STEM Salons that bring together innovative thinkers invested in creating a better STEM world for all. Our first salon focused on the power of social media campaigns like the #XinSTEM initiatives, to accelerate change towards inclusion and leadership by underrepresented communities in STEM: #Change: How Social Media is Accelerating STEM Inclusion[1]. This second paper of the series provides a summary of research-based mentorship practices that have worked at improving the experience in STEM for underrepresented groups.

**OVERVIEW**

Utilizing the 2019 report entitled "The Science of Effective Mentorship in STEMM"[2] published by the National Academies Science Engineering Medicine (NASEM) as a catalyst, the 2030STEM team convened leaders in STEM mentorship initiatives from a variety of fields. The leaders have differing levels of influence and impact for their institutions including university deans, corporate Diversity, Equity, and Inclusion (DEI) executives. They include leading policy researchers that review and strategize the research and findings from the report, , to accelerate and scale effective mentorship for underrepresented groups in STEM.

In understanding how to scale and accelerate, it was important to lay the groundwork of what the most current research represents as the elements of effective mentorship. Two of the researchers serving on the committee that produced the NASEM 2019 report, Dr. Keivan Stassun from Vanderbilt University, and Dr. Christine Pfund of the Wisconsin Center for Education Research at the University of Wisconsin-Madison, led participants in the highlights of the report and how the work is advancing nationally.

With the landscape set, other attendees deepened the discussion of implementation of effective practices currently underway. Dr. Marisela Martinez-Cola of Morehouse College discussed cross-racial mentoring; Dr. David Asai of Howard Hughes Medical Institute (HHMI) summarized the culturally aware mentorship training that is part of the HHMI Gilliam graduate program; Dr. Lou Muglia and Alfred Mays presented on Peer and Near Peer Mentoring Communities supported by the Burroughs Wellcome Foundation; Dr. Preeti Gupta described the social network theory applied to understanding the long-term impact of the Science Research Mentoring Program supported by the American Museum of Natural History; Dr. Erich Jarvis discussed how Rockefeller University approaches their student-led mentorship programs; Dr. Stephanie Danette Preston described the evolution of University Centers of Exemplary Mentoring (UCEM) program; and Dr. Yabebal Fantaye shared how 10Academy's industry-sector mentoring is impacting STEM career pathways in Africa's technology sector.

Three significant themes emerged from these discussions, as a needed focal points for retaining a more diverse STEM workforce, which we will explore in detail in this white paper:
- Cross-Racial Mentoring is needed to maximize the STEM training experience across diverse groups;
- A culture change in STEM mentorship is needed to shift the burden from the mentee to one that is shared among all those in the mentorship ecosystem; and
- Evidence-based mentorship techniques should be utilized as they provide a pathway to better overall experiences for the mentor and mentee.

> *Mentoring 'minority students and/or faculty' requires practicing cultural humility; being willing to take action to tackle racism, discrimination and other inequities; publicly advocating or singing the praises of their scholarly work; and being familiar with the resources available on campus* — Dr. Brad Johnson, professor of psychology, U.S. Naval Academy from Dr. Marisela Martinez-Cola's article "Collectors, Nightlights, and Allies"

**MAKING CROSS-RACIAL MENTORING SUCCEED**

With more students from underrepresented groups entering STEM fields, the dearth of mentors who look like them, due to historical racial bias, leaves students to more often than not rely on mentors from a different racial or cultural background. Given the prevalence of white mentors, we must examine cross-cultural mentoring and develop practices to ensure underrepresented students in STEM receive valuable support and opportunities. Culturally responsive mentoring strategies must become a learned skill set in which mentors, regardless of their race, show interest in and value students' cultural backgrounds and social identities[2].

In Dr. Marisela Martinez-Cola's article titled "*Collectors, Nightlights, and Allies, Oh My!*" published in 2020 in "*Understanding and Dismantling Privilege",* the author defines three roles that white mentors play for students of color and how each differ in positive or negative impact on the mentee[3]. These three roles, which also constitute the title of the article, emerged during a particularly transparent networking session with underrepresented students in an undergraduate fellowship program where the group discussed their experiences with white mentors. The role of white mentors categorized as "Collectors" are genuine in their desire to help, but are misguided and motivated by what has been identified as a "white savior complex." Collectors are likely to trot diverse mentees out to events or ask them to serve on some type of diversity committee. In contrast, "Nightlights", are white mentors who understand the challenges inherent at historically white institutions and can help students of color navigate the unknown and unforeseeable challenges. To their namesake, they figuratively provide light in the dark, unfamiliar places within academia.

Nightlights recognize and acknowledge the existence of systemic racism within the academy, yet, use their privilege, social capital, and cultural capital to reveal the often unspoken rules. Lastly, "Allies" are the mentor category with the most promise. Allies create a two-way relationship and can accept criticism, build trust with consistency and humility and invest in mentees doing their best.[3]

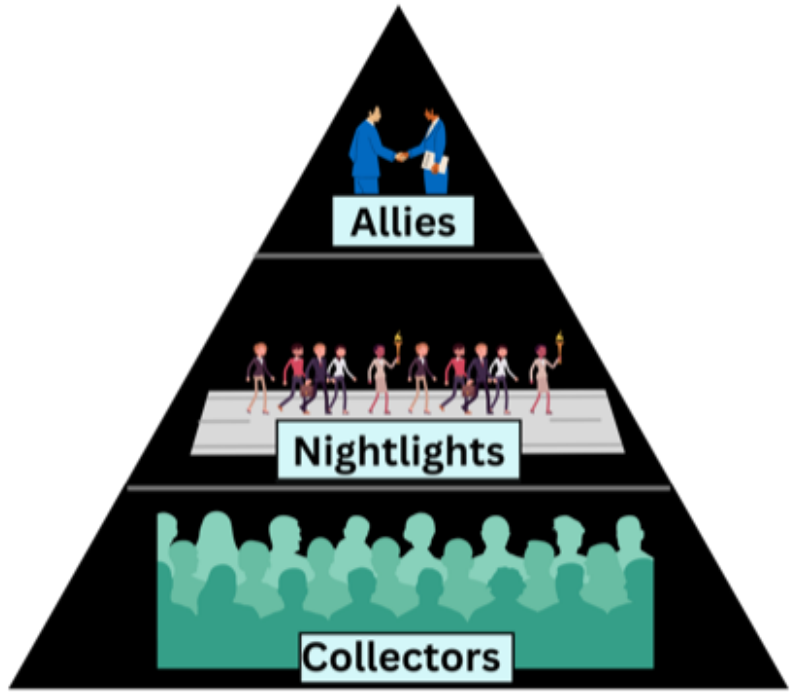

**Figure 1: Categorizing the Role of White Mentors.** Understanding which category a mentor falls into allows for crucial self-reflection for mentors while empowering underrepresented STEM mentees to understand what qualities to seek out in mentors. Adapted from [3].

Dr. Martinez-Cola's research underscores the importance of self-reflection and educating mentors on how best to support the unique needs of underrepresented mentees in STEM. While these categorizations are useful, these are not static roles with clear lines; one person's Ally may be another's Nightlight. Meanwhile, all Allies are also Nightlights, but not all Nightlights are necessarily Allies, and Collectors are never an Ally, but they often sincerely believe that they are an Ally. As increasing representation of underrepresented groups in leadership roles in STEM will take time, it is imperative that we learn to recognize the roles of Collectors, Nightlights, and Allies and alert potential mentees how to mitigate the risks and opportunities of being mentored by these individuals.

Social identity has been proven to impact mentoring relationships as mentees without access to culturally responsive mentoring can experience identity interference, which can result in depression, reduced psychological well-being and lower academic or professional performance[2]. An important step on the journey toward cultural responsiveness is becoming more culturally aware. Culturally aware mentoring refers to a mentor's ability to recognize their own cultural beliefs and judgements as well as similarities and differences between themselves and their mentees and use that information to guide their practice[4]. Culturally aware mentoring is needed as this guides faculty mentors to understand the sources and impact of barriers such as bias on diverse graduate mentees and to improve the training environment for students underrepresented in STEM[5].

> *Like other parts of the scientific enterprise, mentoring needs institutional support, commitment to best practices and innovation, accountability and oversight, and rewards and recognition. Effective mentorship, in other words, requires deliberate and intentional actions at the individual as well as institutional levels. When the scientific establishment fails to train the next generation of scientists in ways that are intentional and effective, both individuals and the academic research enterprise as a whole are shortchanged.* – Beronda L. Montgomery, Fátima Sanchezieto, Maria Lund Dahlberg, "Academic Mentorship Needs a More Scientific Approach"

**CHANGING CULTURE AND SHIFTING THE BURDEN FROM MENTEE TO A SHARED MENTORSHIP ECOSYSTEM**

For far too long, change in inclusion metrics have focused solely on the numbers of individuals underrepresented in STEM who are able to demonstrate career advancement. What is lost in this the focus on this singular metric is the need for a systemic view and for change to occur at all levels of that system. For this to happen, the STEM environment, i.e. culture, must evolve in order to drive effective mentoring relationships that adequately support researchers and professionals from underrepresented groups in culturally responsive ways. The effectiveness of mentorship varies widely due to a number of issues, including lack of training and broader applications of research. A key finding from the NASEM report is that mentorship is as much a science backed by evidence as are other fields of research.

Despite the best intentions, mentorship is not always positive. Ineffective, and sometimes negative, training can occur when mentors are absent, set unrealistic expectations, don't provide clear guidance, or take credit for the mentee's work[2]. An evidence-based approach is needed to empower mentors and mentees with the tools to set clear expectations, engage in regular assessments and participate in mentorship education. This approach will help to set specific constructs of the relationship to make it as effective and impactful for both the mentee and mentor. The importance of accountability of mentors needs to be defined as part of the professional endeavor. While this notion of accountability causes some tension amongst mentors, that accountability is necessary to drive the best outcomes for mentees.

The NASEM report contains a section dedicated to outlining a roadmap for developing a culture of mentorship. This section includes an overview of the most common barriers to organizational change in addition to listing actions, categorized by role, that members of the mentorship ecosystem can take to help foster the development of a culture that supports and values effective mentorship[2].

National programs like the Center for Improvement of Mentored Experiences in Research (CIMER, www.cimerproject.org) and National Research Mentoring Network (NRMN, www.nrmnet.net) are committed to supporting cultural change in mentorship through evidence-based mentorship education and expanding mentoring networks. CIMER investigates approaches for improving mentoring relationships for organizations and institutions. To do so, CIMER develops, implements, and evaluates mentor and mentee training using theoretically-grounded, evidence-based, and culturally-responsive training interventions and investigations. Specifically CIMER:

- Facilitates research mentor and mentee training for mentees and mentors at all career stages;
- Develops and studies new approaches and resources for advancing mentoring relationships;
- Promotes cultural change that values excellence in research mentorship;
- Builds a network of mentors, mentees, and those engaged in enhancing and studying research mentoring relationships; and
- Advances diversity in the research enterprise.

NRMN develops, implements, assesses, and disseminates innovative and effective approaches to engaging, training, and mentoring students, enhancing faculty development, and strengthening institutional research training infrastructure for individuals underrepresented in biomedical research. NRMN's resource center provides a range of offerings for mentors and mentees including a virtual mentorship program, a mentoring network platform and links to many mentorship tools. NRMN's research community studies interventions[1] aimed at advancing the science of mentorship.

---

[1] https://nrmnet.net/nrmn-u01-studies-as-a-part-of-nrmn-phase-ii/

Together, both organizations gather evidence that supports research-based mentorship practices to empower mentors and mentees, and provide guidelines/best practices that can be engaged across the STEM ecosystem.

**EVIDENCE-BASED MENTORING DRIVES BETTER OUTCOMES FOR RETAINING UNDERREPRESENTED GROUPS IN STEM**

To advance the practice of mentorship and ensure its effectiveness we must change how STEM workplace environments are being evaluated. We must move beyond "employee satisfaction surveys" by adding more meaningful benchmarks of impact and action. The programs below reflect how several forward-thinking institutions and organizations are implementing and evaluating formal mentorship education, initiatives that require shared purpose and responsibility between mentors and mentees, and displaying a better understanding of the support students who are underrepresented in STEM require. These examples provide a valuable blueprint of effective mentorship in practice that can be replicated to accelerate and improve mentorship for underrepresented students across STEM sectors.

Culture of Mentorship at the HHMI Gilliam Fellowship

In order to change the culture of science, we have to change its structures and the behaviors of scientists. This notion was the catalyst for how the Gilliam Fellowship for Advanced Study at the HHMI approached the evolution of its program. The Gilliam Fellowship program, which started in 2005 and is intended to prepare students underrepresented in STEM to assume leadership roles in science and science education, was effective at graduating its students. However, many of the alumni chose to leave research and pursue other careers. The HHMI team discovered that, in many cases, the Gilliam Fellows found themselves in a research lab environment that was not prepared to provide culturally aware mentorship.

The HHMI team sought to improve its program, starting with an examination of the structure of the fellowship program. This analysis identified that the responsibility for the success of the Fellows fell mainly on the students; while dissertation advisors play a key role, the program did not emphasize the responsibility of the advisor.

In version 2.0 of the Gilliam Fellowship, which debuted in 2015, structural changes were implemented to: (1) expand who can apply to the program; (2) create a nomination process in which the nominator submits information and data demonstrating the institution's commitment to diversity and how the environment is responsive to DEI; (3) award the grant to both the student and advisor; (4) require all advisors to complete a year-long Mentorship Skills Development course; and (5) add funds to enable the advisor to implement diversity and inclusion activities in their home institution. Paramount in the revised version of the fellowship is the commitment from advisors to be directly involved in all phases of the program. Awarding the grant to both the student and the advisor signals their shared responsibility throughout the application and grant term.

While these structural changes have been significant, the culture of the program would not have evolved without also changing behaviors. Achieving behavior change began with formal training for faculty advisors. Participating mentors are now required to complete a year-long Mentorship Skills Development course created and delivered by the CIMER team. The course aims to help advisors explore their own cultural identities, consider how to connect across cultural differences and reflect on their role in creating a more inclusive research culture. As of September 2022, more than 240 Gilliam advisors have completed the Mentorship Skills Development course. As a result, HHMI has begun to cultivate a network of mentors better equipped to support and advocate for the needs of their mentees while also becoming agents of change themselves. The focus on mentorship development and the direct involvement of advisors in all facets of the program has empowered faculty advisors to become drivers of change in their ecosystem and use mentorship as a lever towards equity and inclusion.[6]

UCEM's Path to Building a Scalable Mentoring Program

Penn State's University Centers of Exemplary Mentoring (UCEM) program, led by Dr. Stephanie Danette Preston was founded in early 2000 to provide graduate students with resources and funding, professional development, community building and advocacy, and mentoring. The Sloan Foundation's Scholarship-funded program, which provides scholars with $40,000 in addition to their tuition assistance along with a wide variety of community building and educational resources, is part of a larger network of scholars from several institutions[4]. The UCEM model is designed to create scale and expand access to mentored support for graduate students underrepresented in STEM. Over the years, the program has evolved to deliver as much value as possible to mentees. When the program began, administrators identified faculty mentors and paired scholars. Throughout the program, some of these mentors proved to be *Collectors* (see Figure 1) while others ultimately were not genuine to the process. As the program evolved, UCEM began providing scholars with a list of vetted mentors for students to choose from and select their own

mentor based on relationships they had cultivated. Additionally, UCEM began facilitating professional development opportunities for mentors to provide formal training while senior scholars also provide mentorship and guidance for scholars just beginning their program. In its 22 years, UCEM scholars have achieved a 95% completion rate with critical support provided by mentoring and relationship-building cultivated and encouraged at all levels–from peer-to-peer and amongst faculty.

Leveraging a Near-Peer Model to Cultivate Interest in STEM

The Burroughs Wellcome Fund (BWF) nurtures a diverse group of leaders in biomedical sciences to improve human health through education and discovery. They offer a variety of diversity mentorship strategies as the organization seeks to cultivate pathways and build communities of diverse scientists. One of its key mentorship initiatives, Promoting Engagement in Science for Underrepresented Ethnic and Racial Minorities (P.E.E.R), partners previous BWF Postdoctoral Diversity Enrichment Program awardees with high school students interested in pursuing a career in STEM. The pilot program is designed to provide formal inclusive mentor training, facilitate near-peer mentorship to high school students, and organize career development activities that engage faculty, P.E.E.R students, parents, and community partners, proactively building interest in STEM. The program is in the early stages with several outcome measures being collected at this time. BWF is also exploring development of competitive grant award programming that may be designed specifically for near-peer mentorship programming between institutions of higher education and high schools.

Improving Retention through Mentorship

For Dr. Erich Jarvis, "hard work combined with talent and opportunity breeds success". While students are responsible for hard work and talent, opportunity is what the community and surrounding ecosystem must supply for students, which is where mentors can make a significant impact. While a faculty member at Duke University, Dr. Jarvis witnessed that highly talented undergraduate and graduate students underrepresented in STEM increasingly shared that they didn't feel a sense of belonging at the primarily white institution, which inhibited their progress and academic success and led some to leaving Duke. Upon receiving this feedback, Duke University School of Medicine created an Office of Biomedical Graduate Diversity, headed by Dr. Sherilynn Black, which approached mentorship programs with an emphasis on supporting mentees to understand and become proactive in identifying and seeking support for their needs and inclusion. The creation of this office had a tangible impact on retention rates for underrepresented graduate students and now serves as a model of change at other levels (undergraduate to faculty) at Duke University and at other institutions.

Dr. Jarvis has since moved to Rockefeller University in 2017 where a similar coalescing of students underrepresented in STEM or of anyone wanting to help is done through a student-led, 2018-launched, Research for Inclusive Science Initiative (RISI). RISI is made up of a diverse group of students that have access to mentors across the faculty and upper administration. In response to the civil unrest following George Floyd and other killings by police, RISI requested and lobbied for the Rockefeller University's first DEI office, and DEI officer, changes in policy for faculty hiring that make it more open to hiring underrepresented students, a lecture series focused on the subject of university, and investment into research on DEI issues. While student mentees require a good degree of resilience to be successful in the sciences, the additional resilience needed from historically underrepresented students to overcome racism and bias, needs to be shifted away from them and to more systemic changes that distributes the burden among stakeholders.

Developing a Sustainable Model for Training and Mentorship in Industry
10Academy is a not-for-profit community-owned initiative that has developed scalable, financially sustainable, and effective STEM mentoring and training programs to build Africa's workforce in AI, Web3, IoT, and Quantum (QIWAI). Their focus is on skill-building for the majority of young Africans who lack access to a college education. Only 10% of the 10 million African young people that enter the workforce annually have access to a college education, and only 1 in 6 college graduates secure related employment. Securing employment for current and future generations of young Africans that want to work is essential. 10Academy's model involves hands-on mentoring where students are partnered with tech industry professionals who could become their colleagues in the future. The 10Academy program identifies trainees during their final year of undergraduate study and provides technical skill development paired with real-world career mentoring in entry-level industry positions. 10Academy then follows the mentees for 3 years as they transition to a senior-level employee.

Demonstrating the impact of this work, a Data Scientist in Ethiopia who completed his training in 2019 and secured a job in 2020 is now supervising six engineers and co-leading his company's AI-led transformation. In total, the intensive skills training and mentored job pairing conducted by 10Academy has led to 90% job placement of the organization's 105 trainees, 38% of whom are women, and salaries 5x that of their peers. While skills training is provided for free, trainees are asked to pay it forward by donating one month of their salary upon being hired to enable free training for the next cohort of students. This near-peer model sustains the program and enables 10Academy to train additional African talent. 10Academy's early career professionals often grow their departments and hire more in-country and in-Africa tech workers, furthering the organization's mission to be a multiplier in building STEM employment opportunities and create a STEM culture among young Africans that drives economic and societal impact.

Making cross-racial mentoring succeed is dependent on changing the culture of mentoring and shifting the burden from mentee to a shared mentorship ecosystem. Systemic culture change is needed to improve outcomes for all early stage researchers, but particularly for researchers from underrepresented groups in STEM who are most susceptible to negative mentorship experiences. As highlighted in various evidence-based mentoring programs, setting specific constructs of the mentoring relationship drives better outcomes for retaining underrepresented groups in STEM.

**RECOMMENDATIONS TO ACCELERATE INCLUSION AT SCALE USING MENTORSHIP**

To accelerate inclusion and leadership for those underrepresented in STEM, we must implement a range of strategies shifting the onus of change from the individual mentee to the STEM environment and culture. With a change in how environments can effectively support researchers and professionals underrepresented in STEM, we have the opportunity to examine systemic changes that make a difference in outcomes for science inclusion and leadership. Below is a synthesis of key recommendations for systemic cultivation and acceleration of effective mentorship of underrepresented groups in STEM:

<u>Institutionalize mentorship education across the STEM ecosystem</u>

Formal training for mentors specifically designed to develop culturally aware and responsive mentoring is needed to ensure mentors are equipped with the knowledge and skills required to support students underrepresented in STEM. Institutions should leverage proven, evidence-based mentorship methodologies to build, scale, and optimize programs. NASEM's "<u>The Science of Effective Mentorship in STEMM</u>" report details the different relationships, structures and behaviors, and institutional cultures that support mentorship and provides insights on effective programs and practices that can be implemented. <u>CIMER</u> and <u>NRMN</u> also provide a range of evidence-based and culturally-responsive resources to develop, implement, assess, and disseminate effective mentoring methodologies using a systems level approach.

Evaluate Environments as a Fundamental Tool for Change

Environmental Climate Evaluations must move beyond "employee satisfaction surveys" to multi-stakeholder, research-based tools that are employed to generate authentic assessments of support systems for labs and other research and professional environments. This approach can then be iterated on to create improved environments for career success for those underrepresented in STEM. Mentorship ecosystems have become an important measure of an environmental shift. New research offers meaningful ways to assess how mentoring relationships are improving career pathways for those underrepresented in STEM[7][8].

Effective environmental evaluation is a prevention measure for institutional leaders to anticipate barriers within their institutions for full inclusion and leadership opportunities for those underrepresented in STEM.

Set Specific Constructs of Accountability and Measure Success Based on Outcomes

While scientific training is often perceived as an apprenticeship, for the practice of mentorship to become more effective, we must engage mentors as well as mentees to set specific constructs of the relationship. Mentorship is a professional commitment that must be done thoughtfully and with tangible outcomes in mind. By balancing accountability with a set of shared goals or purposes that mentees and mentors can agree on, mentoring relationships will be more structured and focused on achieving desired outcomes.

Similar to what has been enacted for effective postdoctoral scholars training in developing and implementing an Individual Development Plan[5], consider asking mentors to develop a detailed mentorship plan to communicate objectives, incentivizing faculty to complete training on effective mentorship, and even tying promotion or tenure criteria to mentorship activities and quality of those relationships[9]. Importantly, this work must be valued, encouraged, and incentivized. Training must be provided in professional development workshops. Assessment of mentorship plan outcomes may be best conducted by department heads, deans, and provosts to identify best-practice of mentorship activities that achieve specific outcomes of mentorship programs. A synthesis of best-practices can then be made accessible to mentors as a framework to identify mentorship plan elements that lead to successful outcomes.

Invest in Cross-Cultural Mentoring Ecosystems

The model of the mentoring ecosystem must be restructured. This requires an investment of time, resources, and funds. Mentors should have dedicated time to develop mentoring plans as an expansion of their training. Resources like professional development programs for creating effective mentoring plans, or for implementation of CIMER and NRMN programs, should be provided by institutions universally and not funded from existing departmental structures. This may require the acquisition of new funds or reallocation of existing capital. Funding the expansion of training and support necessary to change the model of the mentoring ecosystems will require commitment from federal and private funding agencies. Granting and philanthropic organizations can partner with institutions to ensure that funds are allocated toward evidence-based mentor ecosystems to build in financial support for on-going evaluation and sustainability.

In conclusion, instituting these mentoring strategies will remove barriers and accelerate both full inclusion and leadership opportunities for underrepresented groups in STEM.

---

**ACKNOWLEDGMENTS**

2030STEM Inc. gratefully acknowledges funding from the Alfred P. Sloan Foundation (G-2021-16977) and for their inspirational support in our planning year and for our Salon series. 2030STEM also acknowledges all participants of the Mentoring Salon for their thoughtful insight, visionary contributions, and dedication to building a STEM ecosystem that works for all. Holford's work was also supported by National Science Foundation Award (DRL # 2048544).